\documentclass[twocolumn,showpacs,amssymb]{revtex4}

\usepackage{graphicx}
\usepackage{psfig}
\usepackage{dcolumn}
\usepackage{bm}

\def\lessim{\lower.5ex\hbox{$\; \buildrel < \over \sim \;$}}
\def\gtrsim{\lower.5ex\hbox{$\; \buildrel > \over \sim \;$}}

\begin{document} 
\topmargin -0.8cm
\preprint{}

\title{QCD Equations of State and the QGP Liquid Model}

\author{Jean Letessier}
\affiliation{Laboratoire de Physique Th\'eorique et Hautes Energies\\
Universit\'e Paris 7, 2 place Jussieu, F--75251 Cedex 05.
}
\author{Johann Rafelski}
\affiliation{Department of Physics, University of Arizona, Tucson, Arizona, 85721, USA}

\date{January 14, 2003}

\begin{abstract}
Recent advances in the study of equations of state of thermal  lattice
Quantum Chromodynamics obtained at non-zero baryon density  
allow validation   of the 
quark-gluon plasma (QGP) liquid model equations of state (EoS). We study here 
the properties of the  QGP-EoS  near to the phase transformation 
boundary at finite baryon density and show a close agreement with
the lattice results. 
\end{abstract}

\maketitle

%\section{Introduction}
%%%%%%%%%%%%%%%%%%%%%%%%
The ab-initio exploration of Quantum Chromodynamics (QCD) on the lattice 
at finite temperature and baryon density has seen 
rapid recent progress~\cite{Fod02,Fod02b,All02}. 
It is already known  that for vanishing  baryon density  lattice  results
can be described in terms of ideal quantum gases allowing only for lowest
order perturbative interactions, provided that a non-perturbative
temperature dependence of the coupling constant is introduced,
along with vacuum latent heat~\cite{Ham00}.  We will
%here 
refine this quark-gluon plasma model,  and
 demonstrate that it agrees extremely well with the finite 
baryon density lattice results,
except in a region of temperature in vicinity of the phase 
transition/transformation domain.

A strong motivation to have an understanding of the thermal QCD
lattice results in 
terms of a set of more intuitive degrees of freedom originates in the need to 
treat  a fast evolving system 
created in relativistic heavy ion collisions. At the onset of 
the QCD thermal matter formation,
 {\it e.g.}, at $\tau_0\simeq 0.5$\,fm, one cannot expect a full 
chemical equilibrium to have been reached, {\it i.e.}, the 
phase space occupancy of quark and/or gluon gases has not 
approached unity. In a model as we will present these
situations can be easily incorporated, allowing study of
initial temperature, phase space occupancy evolution,
and other physical conditions
in the realistic environment of 
high energy relativistic heavy ion collisions. 

More generally,
a handy model of QCD matter equations of state employing established
physical concepts allows to study many physical properties which are
only difficult to infer from the numerical lattice results. But perhaps 
the most important question we pursue is if one can
 already in the temperature domain explored by present 
day experiments describe thermal QCD matter in terms of nearly 
free quark and gluon degrees of freedom. If this is 
confirmed, we would be able to verify if  the physical system 
formed in these reactions is  the deconfined quark-gluon plasma. 

\begin{figure}[!b]
\vspace*{-.16cm}
\hspace*{0.cm}\psfig{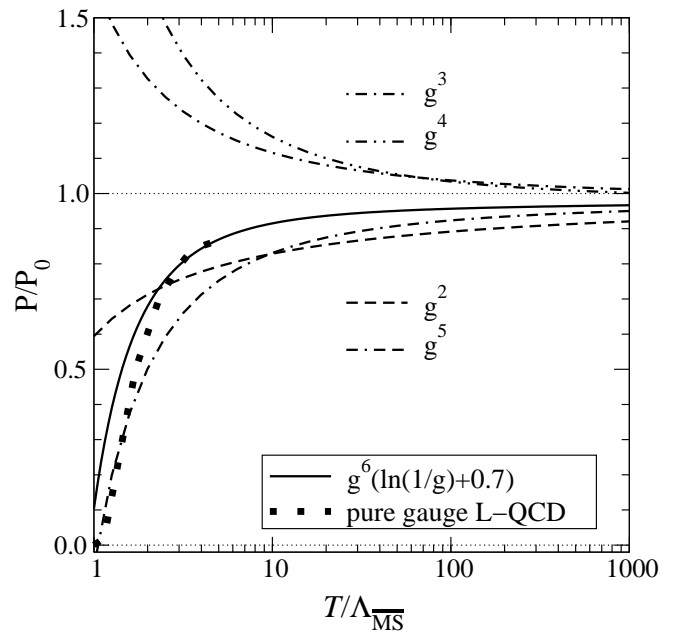}
\vspace*{-.4cm}
\caption{ %\small
Ratio of the pure gauge pressure  with Stefan-Boltzmann pressure $P/P_0$
as function of temperature $T$ in units of renormalization scale 
$\Lambda_{\overline{\rm MS}}$, results of  Ref.~\protect\cite{Kaj02}.
\label{Kajang6ln}
}
\end{figure}

A systematic perturbative expansion  within the framework of 
thermal field theory of the interacting quark-gluon gas 
converges poorly~\cite{Kaj02}. The situation is illustrated for 
the case of a pure gauge SU(3) case in figure~\ref{Kajang6ln}, where
we see that the ratio of the computed perturbative  pressure  $P$ to 
the Stefan-Boltzmann pressure $P_0$ 
is oscillating around unity depending on the order $n$ of the expansion 
in the coupling constant $g$ considered. Convergence seems to occur only
in the asymptotic freedom
limit of relatively high temperature $T$, beyond the experimental 
reach. Better agreement with the 
lattice results (solid dots) can be achieved (solid line) 
at a relatively low $T$, here
expressing the temperature in units of the renormalization 
scale $\Lambda_{\overline{\rm MS}}$,  provided that  the  unknown 
relative coefficient of the $g^6$ term is optimized to the value 0.7\,.
%%%%%%%%%%%%%%%%%%%%%%%%%%%%%%%  Figure 0
%%%%%%%%%%%%%%%%%%%%%%%%%%%%%%%%%%%%%%%%%%%%

We advocate a very different approach, combining
the perturbative result with key nonperturbative
features. 
Our  QGP liquid model arises from the empirical observation that the 
lowest order thermal  perturbation contribution evaluation of the QCD matter
properties, combined with a non-perturbative temperature dependent strong 
coupling constant agrees with the QCD thermal lattice results, once 
the other often used  non-perturbative effect, {\it e.g.} bag constant, has been
incorporated. 

The QGP-liquid partition 
function is assumed to have the form,
\begin{eqnarray}
\label{ZQGPL}
&&\frac{T}{V}\ln{\cal Z}_{\mathrm QGP}
\equiv P_{\mathrm QGP}=
-{\cal B}+\frac{8}{45\pi^2}c_1(\pi T)^4  \\[0.3cm]  
&&+
\sum_{i=q,s}\frac{n_i}{15\pi^2}
\left[\frac{7}{4}c_2(\pi T)^4+\frac{15}{2}c_3\left(
\mu_i^2(\pi T)^2 + \frac{1}{2}\mu_i^4
\right)\right]
\nonumber
%\\[0.3cm]  
%&&+
%\frac{n_{\mathrm s}}{15\pi^2}
%\left[\frac{7}{4}c_2(\pi T)^4+\frac{15}{2}c_3\left(
%\mu_{\mathrm s}^2(\pi T)^2 + \frac{1}{2}\mu_{\mathrm s}^4
%\right)\right],\nonumber
\end{eqnarray}
 where $n_q=2$, $n_s=1$, $\mu_s=0$ and:
\begin{eqnarray}
\label{ICZQGP}
c_1&=&1-\frac{15\alpha_s}{4\pi}\,,\nonumber\\[0.1cm]  
c_2&=&1-\frac{50\alpha_s}{21\pi}\,,\qquad
c_3=1-\frac{2\alpha_s}{\pi}\,.
\end{eqnarray}
We recall that $\mu_b=3\mu_q$ and $\lambda_q=e^{\mu_q/T}$. A value 
${\cal B}=(0.211\,\mbox{GeV})^4$ fits best the lattice results 
we consider here.

The temperature dependence $\alpha_s(T)$ is
obtained from $\alpha_s(x)$ setting the energy scale $x$
to be:
\begin{equation}
\label{runalTmu}
x=2\pi \beta^{-1}\sqrt{1+\frac{1}{\pi^2}\ln^2\lambda_{\mathrm q}}
=2\sqrt{(\pi T)^2+\mu_{\mathrm q}^2}\,.
\end{equation}
$\alpha_s(x)$ is obtained integrating the renormalization group equation,
incorporating physical thresholds for heavy flavor. Use of semi-analytical 
formulas with fixed active number of quarks introduces
an unacceptable error in the value of $\alpha_s$ and 
has lead  to false conclusions in some earlier work. We evaluate:
\begin{equation}
\label{dalfa2loop}
x \frac{\partial \alpha_s}{\partial x}=
-b_0\alpha_s^2-b_1\alpha_s^3+\ldots \equiv \beta^{\mbox{\scriptsize pert}}_2\,.
\end{equation}
$\beta^{\mbox{\scriptsize pert}}_2$ is
the beta-function of the renormalization group 
in two loop approximation, and 
$$b_0=\frac{11-2n_{\mathrm f}/3}{2\pi}\,,\quad 
   b_1=\frac{51-19n_{\mathrm f}/3}{4\pi^2}\,.$$ 
$\beta^{\mbox{\scriptsize pert}}_2$
does not depend on the renormalization scheme,
and solutions of Eq.\,(\ref{dalfa2loop}) differ from higher 
order renormalization scheme
dependent results  by less than the error introduced by the experimental 
uncertainty in the measured value of $\alpha_s(\mu=M_Z)=0.118+0.001-0.0016$,
used as the initial value in numerical integration of Eq.\,(\ref{dalfa2loop}).
We note for $\mu_{\mathrm q}\to 0$  the empirical form:
\begin{equation}
\alpha_s(T/T_c)\simeq {0.47\over {1+0.72\ln(T/T_c)}}\,,\quad
   T_c=172\,\mbox{MeV}\,.
\end{equation}

When we explore the effect of the finite 
quark masses, we introduce the correction 
$\delta_m \ln Z^q_{\rm QGP}$
to the partition  function,
\begin{eqnarray}%{equation}
\frac{1}{T^3V}\delta_m \ln Z^q_{\rm QGP}&=&\nonumber 
n_i{m_i^3\over T^3}{3\over \pi^2}\\[0.3cm] \label{delZ}
&&\hspace{-3.2cm} \times \int_0^\infty\!\! dx x^2\!
\left[\ln\!\left(\!\frac{1+\lambda_i e^{-\sqrt{1+x^2}{m_i\over T}}}
    {1+\lambda_i e^{- x {m_i\over T}}}\!\right)\!
   +\!\left(\!\lambda_i\to {1\over\lambda_i}\!\right)\right] ,
\label{dPmq}
\end{eqnarray}%{equation}
where $i=q,s$. We use $m_s=170$\,MeV. Note that we
did not allow for the QCD-thermal $\alpha_s$ correction 
in $\delta_m\ln Z^q_{\rm QGP}$. When we  compare withe lattice results, we use 
the values $m_q=65$\,MeV and $m_s=135$\,MeV, as reported to have 
been used in  lattice simulations we compare with. 
Strange quarks enter below only when we consider the absolute 
pressure. Similarly, when we explore how  an effective gluon mass
alters the pressure,  we consider a correction 
\begin{equation}
\label{dPmG}
\frac{\delta_m \ln Z^G_{\rm QGP}}{T^3V}=
-{m_G^3\over T^3}{8\over \pi^2}
\int_0^\infty\!\!\!  dx x^2\ln\!\left(\!\frac{1- e^{-\sqrt{1+x^2}{m_G\over T}}}
    {1- e^{- x {m_G\over T}}}\right)\!.
\end{equation}
 We  need to introduce a finite thermal glue
mass $m_G\simeq 0.2$\,GeV in order to reach for $T>T_c$ a line width 
agreement with lattice results. This is not the thermal gluon
mass which, as we have discussed elsewhere, can be used as an alternative
way to express the perturbative QCD effect~\cite{Raf02}.

In order to be able to consider our results in currently ongoing 
experiments, it is important that 
the QGP-liquid equations of state are verified at chemical
potential which is relatively small, up to $\mu_b\simeq 100$ MeV at initial 
conditions. However, we extend this  study  to all values considered
in the lattice simulation~\cite{Fod02b}, $\mu_b=100, 210, 330, 410$ and 530 MeV,
which aside of RHIC, also encompasses the physical reach of CERN-SPS experiments.
Our objective is to test if for $T>1.5T_c$ the model and lattice-`experiment' agree,
while near to the critical temperature domain significant modifications
due to the non-perturbative features of QCD must be expected. 

Physical properties  we consider will be rendered dimensionless by considering a 
suitable ratio with either $T^4$ (pressure $P$) or $T^3$ (baryon density $n_b$). 
To asses if the liquid model has the right density dependence 
we consider in turn the  pressure $P$ modification at finite 
baryon density, $\Delta P\equiv P(T,\mu_b)-P(T,\mu_b=0)$,
the baryon density $n_b$, the pressure $P(T,\mu_b=0)$ and 
finally $\epsilon(T,\mu_b)-3P(T,\mu_b)$, which vanishes
for ideal gases. In the following figures all `experimental' 
points are taken directly  from the postscript 
format of the figures of Ref.\cite{Fod02b}, into which 
we add our results.

The reader should  note that  for $T<T_c$ 
equations of state with hadrons must be employed. Hence we do not here
include results for this domain of lattice results.
The study  of this  confined phase domain involves also modeling of excluded volume
effect,  and/or within the bootstrap model, the understanding of the 
singularity of the hadron mass spectrum.  Given the availability of
lattice results we hope to return to 
discuss these intricate matters in near future. 

\begin{figure}[!t]
\vspace*{-.2cm}
\hspace*{-0.4cm}\psfig{width=9.cm,height=11cm,clip=,figure=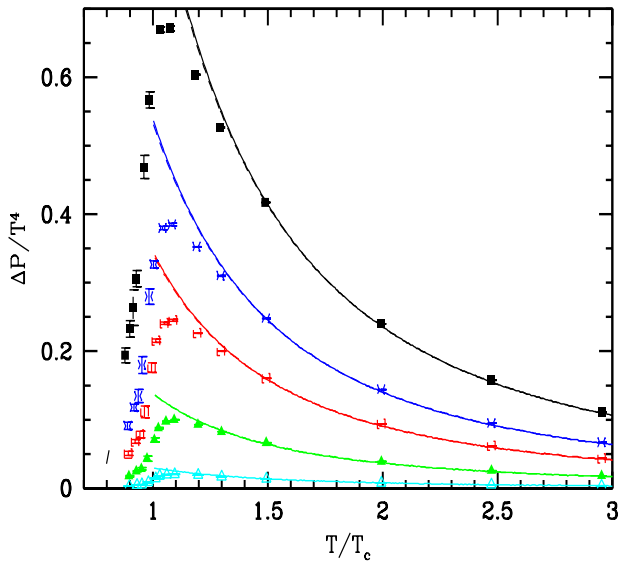}
\vspace*{-3.2cm}
\caption{ %\small
 $\Delta P\equiv P(T,\mu_b)-P(T,\mu_b=0)$ normalized by $T^4$  as function
of $T/T_c$ for $\mu_b=100$, 210, 330, 410 and 530 MeV from bottom to top. Data 
points from  ref.~\protect\cite{Fod02b}, solid lines massless liquid of 
quarks. Dashed (and mostly invisible) results with finite mass 
correction applied for $m_q=65$ MeV as used in lattice data.
\label{DelPresFod}
}
\end{figure}

The change on $\Delta P$ at finite baryochemical
potential is shown in figure \ref{DelPresFod}. We note  that the 
agreement with the lattice results at large $T$ is very satisfactory. 
This result  depends  on the empirical choice
made for the  dependence on chemical potential
of the scale of $\alpha_s$, see Eq.\,(\ref{runalTmu}). 
Had we omitted $\mu_q$, or doubled the coefficient
of $\mu_q$, there would be  a well visible deviation 
in figure \ref{DelPresFod}. Thus the empirical choice of the Matsubara
frequency as the combination of temperature and chemical potential
is  qualitatively verified 
by lattice results.  For small chemical potential
$\mu_b=100$ MeV, we see that there is agreement for $T\ge 1.1T_c$, at a
baryochemical potential $\mu_b=530$ MeV, agreement with lattice data is
assured for $T\ge 1.5T_c$. 
The practically invisible dashed lines (hidden mostly under solid lines)
show the here negligible effect of finite quark masses. By definition,
$\Delta P$ depends only on light quark degrees of freedom. The agreement 
we see is thus not testing strange quark, or glue behavior. 
%%%%%%%%%%%%%%%%%%%%%%%%%%%%%%%  Figure 1
%%%%%%%%%%%%%%%%%%%%%%%%%%%%%%%%%%%%%%%%%%%%

%%%%%%%%%%%%%%%%%%%%%%%%%%%%%%%  Figure 2
\begin{figure}[!t]
\vspace*{-.2cm}
\hspace*{-0.4cm}\psfig{width=9.cm,height=11cm,clip=,figure=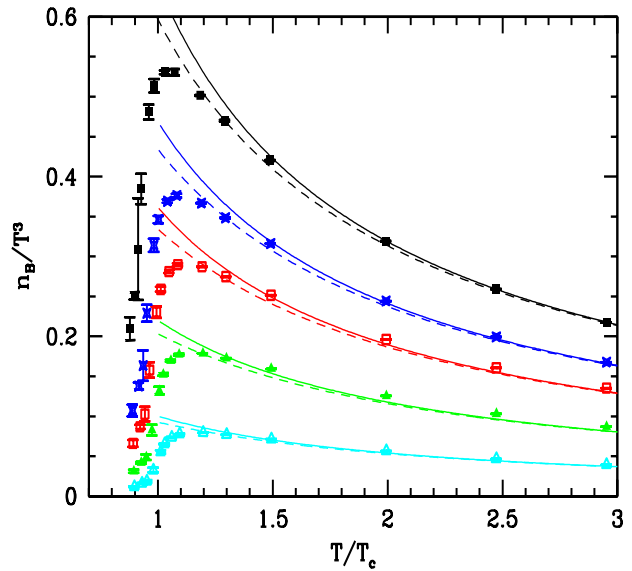}
\vspace*{-3.2cm}
\caption{ %\small
Baryon density $n_{\rm B}$  normalized by $T^3$  as function
of $T/T_c$ for $\mu_b=100$, 210, 330, 410 and 530 MeV from bottom to top. Data 
points from  ref.~\protect\cite{Fod02b}, solid lines massless liquid of 
quarks. Dashed lines: allowance is made for $m_q=65$ MeV as is used to
obtain the lattice data.
\label{BdensFod}
}
\end{figure}
%%%%%%%%%%%%%%%%%%%%%%%%%%%%%%%%%%%%%%%%%%%%

A similarly remarkable agreement is obtained for the baryon density,
shown in figure \ref{BdensFod}. Here a finite quark mass used in 
lattice simulations becomes visible (dashed lines).  Once we allow for it, we see 
agreement within the lattice data error for $T\ge 1.2T_c$
for all baryochemical potentials. For very small baryon density we can expect
agreement down to near the critical temperature. 

Since both the pressure difference $\Delta P$ and baryon density $n_B$ 
show satisfactory 
behavior as function of chemical potential, it is expected that
the change of energy density, and entropy, with chemical potential is 
well described, both being derived of $\Delta P$ with respect
to $\beta=1/T$ and $T$ and  suitable linear combination with $n_B$. 
However, in order to fully certify  the liquid model, we need now  to 
fine-tune the behavior of pressure $P$ at zero chemical potential.

%%%%%%%%%%%%%%%%%%%%%%%%%%%%%%%  Figure 3
\begin{figure}[!b]
\vspace*{-.2cm}
\hspace*{-0.4cm}\psfig{width=9.cm,height=11cm,clip=,figure=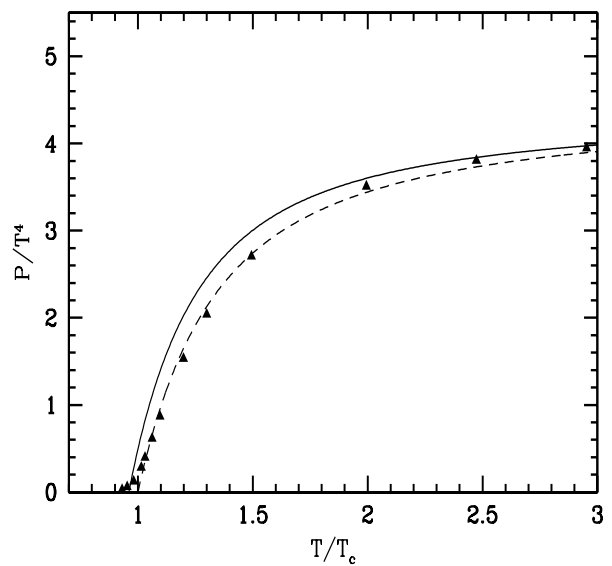}
\vspace*{-3.2cm}
\caption{ %\small
 The pressure $P(T,\mu_b=0)$ normalized by $T^4$  as function
of $T/T_c$. Data 
points from  ref.~\protect\cite{Fod02b}. Solid lines:  
massless gluons with ${\cal B}=(0.211\,\mbox{GeV})^4$.
Dashed line allows for a finite mass $m_G=200$ MeV.
\label{PresFod}
}
\end{figure}
%%%%%%%%%%%%%%%%%%%%%%%%%%%%%%%%%%%%%%%%%%%%

%%%%%%%%%%%%%%%%%%%%%%%%%%%%%%%  Figure 4
\begin{figure}[!t]
\vspace*{-.35cm}
\hspace*{-0.4cm}\psfig{width=9.cm,height=11cm,clip=,figure=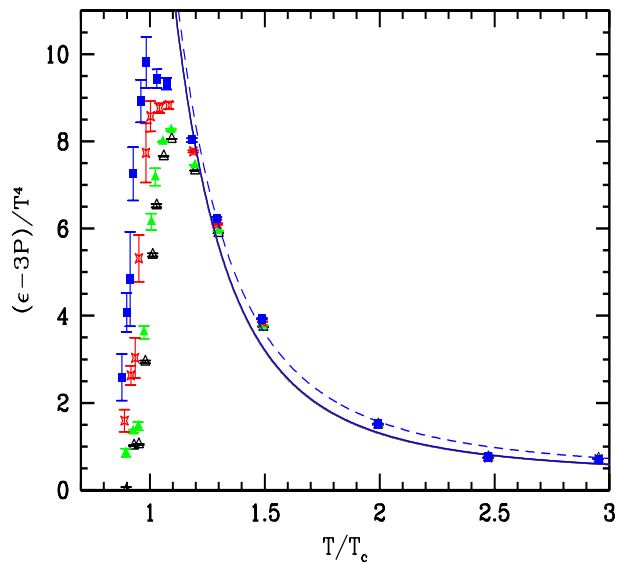}
\vspace*{-3.2cm}
\caption{ %\small
 $(\epsilon-3P)/T^4$  as function
of $T/T_c$ for $\mu_b=0$, 210, 410 and 530 MeV, for 
 ${\cal B}=(0.211\,\mbox{GeV})^4$ and $T_c=173\,$MeV. Data 
points from  ref.~\protect\cite{Fod02b}. Solid lines: massless gluons.
Dashed lines: allowance is made for $m_G=200$ MeV. All chemical potential
lines coincide within line-width.
\label{E3PFod}
}
\end{figure}
%%%%%%%%%%%%%%%%%%%%%%%%%%%%%%%%%%%%%%%%%%%%

In figure \ref{PresFod}, we show the pressure 
$P(T,\mu_b=0)/T^4$ as function of temperature $T/T_c$. 
We find that  there is some difference in the lattice result shown
in   \cite{Fod02b} and earlier work \cite{Kar00}, which practically agreed 
with our solid line result in figure \ref{PresFod}. To obtain the QGP
model agreement with the more recent  lattice 
data  \cite{Fod02b} we need to:\\
a) change the value of the 
 Bag constant, from our earlier value $(0.195\,\mbox{GeV})^4$ to
${\cal B}=(0.211\,\mbox{GeV})^4$, in response to somewhat higher
value of $T_c$, see solid line in  figure \ref{PresFod},\\ %  and,\\
 b) introduce an effective gluon mass $m_G=0.2$\,GeV
in order to reduce the number of effective degrees of freedom as is shown 
by the dashed line in figure \ref{PresFod}.\\
 This procedure does indeed 
produce the expected agreement within the line-width.

We do not have a lattice
entropy figure to compare with, but we are assured of a  valid result
by our ability to reproduce the shape of the pressure functions, see
figures~\ref{DelPresFod} and \ref{PresFod}. A sensitive test of this 
assertion is obtained considering $(\epsilon-3P)/T^4$ as function
of $T/T_c$ in figure \ref{E3PFod}. All finite chemical curves we
plot coincide and hence only one is visible in the diagram. 
This agreement with lattice results for $T>1.15 T_c$  confirms 
that we have obtained a remarkably  
precise representation of the behavior of equations of state
of deconfined QCD matter, except  in direct
vicinity of the critical temperature.

We have shown here that QCD equations of state at nonzero baryon density 
for $T>1.5T_c$, but even  very near to $T=T_c$ for
small chemical potentials, behave in a way expected for the quark-gluon
plasma phase. The remarkable agreement has been made possible by 
introduction of the first order thermal interactions which comprise
precise nonperturbative thermal coupling $\alpha_s(T,\mu_q)$ and the vacuum
latent heat $\cal B$. In the glue sector, effective gluon mass 
$m_G\simeq 200$\,MeV is also required to obtain exactly the 
expected number of degrees of freedom, required to model the 
absolute magnitude of the pressure to within line-width.
It is important to realize that even without such detailed
refinements which may indeed change again as the lattice
results evolve, the approach we advocate does not suffer from the
convergence problem illustrated in figure  \ref{Kajang6ln} and 
it provides a very good and natural description of the wealth of
the lattice results.

We conclude that thermal lattice QCD matter
behaves just like  quark gluon plasma, and thus a
naive use of the QGP model is appropriate, provided that
suitable  coupling strength $\alpha_s(T,\mu_q)$
and vacuum bag constant $\cal B$ is introduced. 
Moreover, considering  that at RHIC the 
following initial conditions  have been reached~\cite{Raf03}, 
$T>1.5T_c\simeq 260$\,MeV  for $\lambda_q=1.09$, corresponding  to the 
initial baryochemical potential 
$\mu_b=3T\ln \lambda_q\simeq 45$ MeV,
the results presented confirm that the QGP state is 
established in these reactions. Moreover, we are 
reassured that we can explore  in detail 
the RHIC initial conditions, allowing for chemical composition 
dynamics. This  should lead us to understanding of QGP 
properties and conditions in RHIC reactions.

{\it Acknowledgments:\/} 
Work supported in part by a grant from the U.S. Department of
Energy,  DE-FG03-95ER40937\,. LPTHE, Univ.\,Paris 6 et 7 is:
Unit\'e mixte de Recherche du CNRS, UMR7589.\\

%\newpage
%%%%%%%%%%%%%%%%%%%%%%%%%%%%%%%%%%%%%%%%%

%%%%%%%%%%%%%%%%%%%%%%%%%%%%%%%%%%%%%%%%%
\vskip 0.3cm
%%%%%%%%%%%%%%%%%%%%%%%%%%%%%%%%%%%%%%%%%
%\begin{references}


\begin{thebibliography}{19}
\providecommand{\bibinfo}[2]{#2}

\bibitem{Fod02b}
\bibinfo{author}{Z. Fodor, S.D. Katz, and K.K. Szabo},
``The QCD equation of state at nonzero densities: Lattice result'',
e-Print Archive: hep-lat/0208078.

\bibitem{Fod02}
\bibinfo{author}{Z. Fodor, and S.D. Katz},
%LATTICE DETERMINATION OF THE CRITICAL POINT OF QCD AT FINITE T AND MU. 
JHEP 0203:014, (2002).

\bibitem{All02}
\bibinfo{author}{C.R.~Allton, S.~Ejiri, S.J.~Hands, O.~Kaczmarek, F.~Karsch, 
E.~Laermann, C.~Schmidt, and L.~Scorzato,}
%The QCD thermal phase transition in the presence of a small chemical potential
  \bibinfo{journal}{Phys. Rev. D} \textbf{\bibinfo{volume}{66}},
  \bibinfo{pages}{074507} (\bibinfo{year}{2002}).

\bibitem{Ham00}
\bibinfo{author}{S.~Hamieh, J.~Letessier, and J.~Rafelski},
  \bibinfo{journal}{Phys. Rev. C} \textbf{\bibinfo{volume}{62}},
  \bibinfo{pages}{064901} (\bibinfo{year}{2000}).

\bibitem{Kaj02}
\bibinfo{author}{K. Kajantie, M. Laine, K. Rummukainen, and Y. Schroder,}
``The pressure of hot QCD up to $g^6 ln(1/g)$'',
e-Print Archive: hep-ph/0211321, {\it and references therein}.

\bibitem{Raf02}
\bibinfo{author}{J. Rafelski, and J. Letessier,}
%Strangeness and Statistical QCD
  \bibinfo{journal}{Nucl. Phys. A} \textbf{\bibinfo{volume}{702}}
  \bibinfo{pages}{304}  (\bibinfo{year}{2002}).


\bibitem{Kar00}
\bibinfo{author}{F.~Karsch, E.~Laermann, and A.~Peikert,}
  \bibinfo{journal}{Phys. Lett. B} \textbf{\bibinfo{volume}{478}}
  \bibinfo{pages}{447}  (\bibinfo{year}{2000}).

\bibitem{Raf03}
\bibinfo{author}{J. Rafelski, and J. Letessier}, 
``Testing limits of statistical hadronization'',
e-print Archive: nucl-th/0209084, to appear in {\it Nucl. Phys.}, A, (2003),
proceedings of  Quark Matter Conference
held 18-24 July 2002, in  Nantes, France.

%\end{references}
\end{thebibliography}
\end{document}